\documentclass[10pt, a4paper]{article}

\usepackage{lrec2022} 

\usepackage{natbib}
\usepackage{multibib}
\makeatletter
\def\@mb@citenamelist{cite,citep,citet,citealp,citealt,citepalias,citetalias}
\makeatother
\newcites{languageresource}{~}

\usepackage{graphicx}
\usepackage{tabularx}
\usepackage{booktabs}
 \usepackage{array,multirow}
\usepackage{verbatim}
\usepackage{subcaption}
\usepackage{soul}
\usepackage{titlesec}
\titleformat{\section}{\normalfont\large\bf\center}{\thesection.}{1em}{}
\titleformat{\section}{\normalfont\large\bfseries\center}{\thesection.}{1em}{}
\titleformat{\subsection}{\normalfont\SmallTitleFont\bfseries\raggedright}{\thesubsection.}{1em}{}
\titleformat{\subsubsection}{\normalfont\normalsize\bfseries\raggedright}{\thesubsubsection.}{1em}{}
\renewcommand\thesection{\arabic{section}}
\renewcommand\thesubsection{\thesection.\arabic{subsection}}
\renewcommand\thesubsubsection{\thesubsection.\arabic{subsubsection}}

\usepackage{xcolor}
\usepackage{hyperref}
 \definecolor{darkblue}{rgb}{0, 0, 0.5}
  \hypersetup{colorlinks=true, citecolor=darkblue, linkcolor=darkblue, urlcolor=darkblue}

\usepackage{colortbl}

\definecolor{red}{rgb}{255,0,0}

\usepackage{xstring}

\title{SCOUT: A Situated and Multi-Modal Human-Robot \\Dialogue Corpus}

\name{Stephanie M. Lukin\textsuperscript{\rm 1}, Claire Bonial\textsuperscript{\rm 1}, Matthew Marge\textsuperscript{\rm 2}, Taylor Hudson\textsuperscript{\rm 3}, 
  \\
  \bfseries\large Cory J. Hayes\textsuperscript{\rm 1}, Kimberly A. Pollard\textsuperscript{\rm 1}, Anthony Baker\textsuperscript{\rm 1}, Ashley N. Foots\textsuperscript{\rm 1}, \\ 
  \bfseries\large Ron Artstein\textsuperscript{\rm 4}, Felix Gervits\textsuperscript{\rm 1}, Mitchell Abrams\textsuperscript{\rm 1}, Cassidy Henry\textsuperscript{\rm 1}, \\ 
  \bfseries\large  Lucia Donatelli\textsuperscript{\rm 5}, Anton Leuski\textsuperscript{\rm 4}, Susan G. Hill\textsuperscript{\rm 1}, \\
  \bfseries\large David Traum\textsuperscript{\rm 4} and Clare R. Voss\textsuperscript{\rm 1}} 

\address{
\textsuperscript{\rm 1}DEVCOM Army Research Laboratory,
\textsuperscript{\rm 2}DARPA,
\textsuperscript{\rm 3}Oak Ridge Associated Universities,\\
\textsuperscript{\rm 4}USC Institute for Creative Technologies,
\textsuperscript{\rm 5}Vrije Universiteit\\
         stephanie.m.lukin.civ@army.mil\\}

\abstract{
We introduce the Situated Corpus Of Understanding Transactions (SCOUT), a multi-modal collection of human-robot dialogue in the task domain of collaborative exploration. The corpus was constructed from multiple Wizard-of-Oz experiments where human participants gave verbal instructions to a remotely-located robot to move and gather information about its surroundings. SCOUT contains 89,056 utterances and 310,095 words from 278 dialogues averaging 320 utterances per dialogue. The dialogues are aligned with the multi-modal data streams available during the experiments: 5,785 images and 30 maps. The corpus has been annotated with Abstract Meaning Representation and Dialogue-AMR to identify the speaker's intent and meaning within an utterance, and with Transactional Units and Relations to track relationships between utterances to reveal patterns of the Dialogue Structure. We describe how the corpus and its annotations have been used to develop autonomous human-robot systems and enable research in open questions of how humans speak to robots. We release this corpus to accelerate progress in autonomous, situated, human-robot dialogue, especially in the context of navigation tasks where details about the environment need to be discovered. 
\\ \newline  }

\begin{document}

\maketitleabstract

\section{Introduction}

For robots to team effectively with humans, a critical
capability will be to use forms of natural communication like
language. Moreover, these interactions must be bi-directional, as robots
will need to provide status updates and ask for and receive clarification or help from
teammates in challenging situations. Finally, the interactions must be situated in knowledge about the environment that the robot inhabits.  
In order to study these forms of communication and accelerate progress in the development of autonomous robot dialogue systems, collections of data should (1) focus on unconstrained, \textit{robot-directed} dialogue in contrast to traditional human-human dialogue,%
\footnote{Humans instruct robots differently compared to instructing other humans~\citep{mavridis2015review,marge2020let,MARGE2022101255}.} 
(2) exhibit natural diversity of communication strategies inherent in situated dialogue, and (3) be organized into a format that can be quickly labeled and used for training an autonomous dialogue system.

\begin{figure*}[h!]
    \centering
    \includegraphics[width=0.8\textwidth]{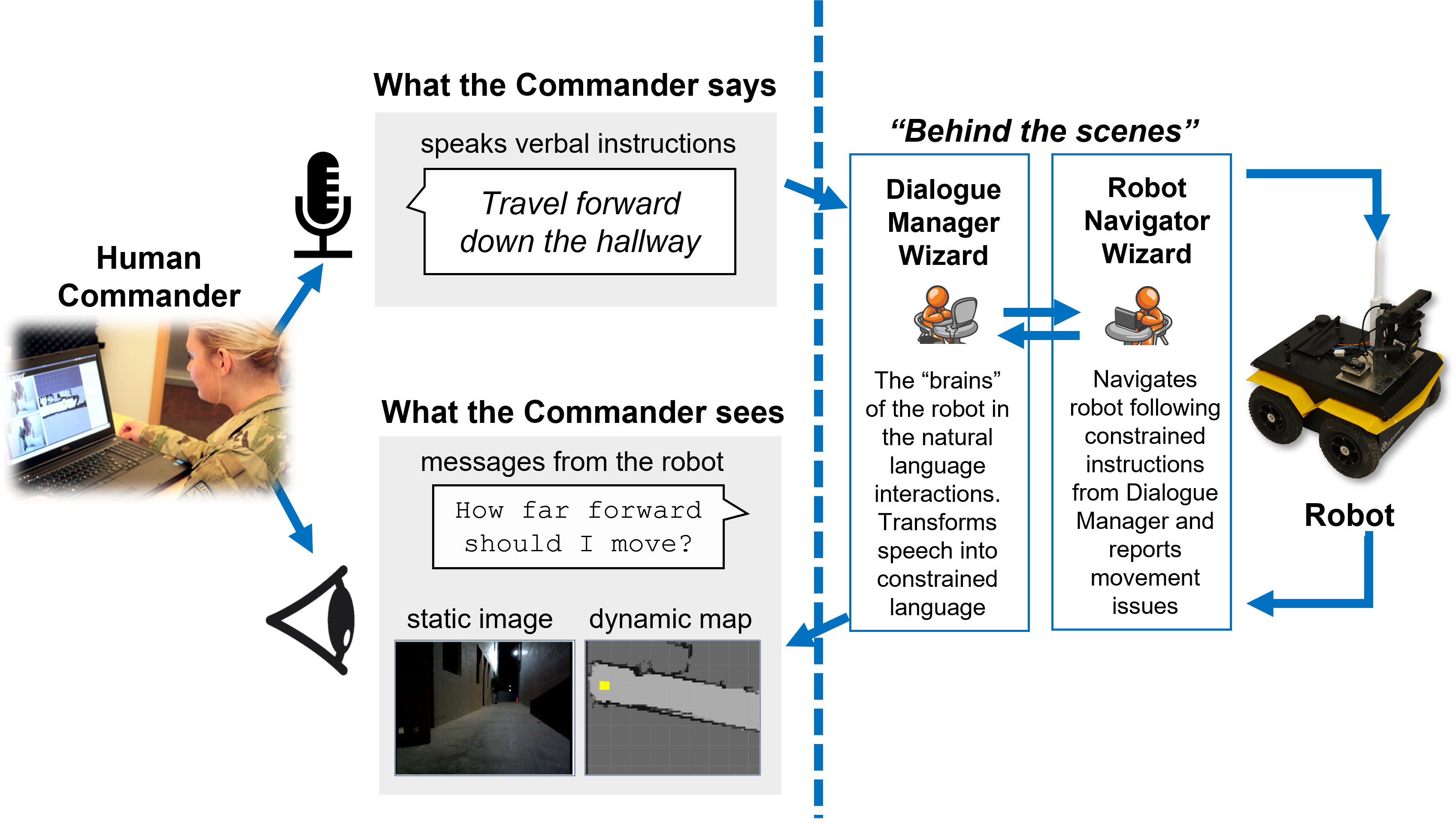}
    \caption{\label{fig:method} Wizard-of-Oz data collection design}
\end{figure*}

In this paper, we present {\bf SCOUT, the Situated Corpus Of Understanding Transactions}, a multi-modal, human-robot dialogue corpus that meets these data criteria. SCOUT is a collection of human-robot dialogues within the task domain of collaborative navigation between a human and a remotely-located robot. Human participants assumed the role of {\it Commander} and collaborated with a remotely-located robot to explore and assess the robot's environment and locate objects of interest. To progress through the task, Commanders relied on a combination of overhead maps generated from streaming LIDAR (LIght Detection And Ranging) sensors, pictures from the robot's camera upon request, and text messages. 
Commanders spoke freely to the robot and were given no restrictions on how they formulated their language, allowing them to follow their natural tendencies for speaking to a robot when completing this task. The dialogues were collected in a Wizard-of-Oz (WoZ) experimental paradigm, wherein the robot's autonomy was controlled by two ``wizard'' experimenters: a \textit{Dialogue Manager} to ground instructions to the robot's surroundings and select dialogue behaviors, and a \textit{Robot Navigator} to move the robot and provide status updates (see Fig.~\ref{fig:method}). 

Our construction of SCOUT brings together for the first time all the data streams and modalities collected in 
this task domain. SCOUT contains 89,056 utterances and 310,095 words from 278 dialogues lasting about 20-minutes each with an average of 320 utterances per dialogue. 5,785 images were taken across the corpus and are linked to the moment they were taken in the dialogue. 
Thirty LIDAR maps are included with annotations that reflect the spaces the robot scanned in the environment during the dialogue (some of the maps  include only subsets of the whole area).

The dialogues exhibit a situated and multi-modal nature with linguistic references to space, distance, and the physical world, providing for many opportunities to study dialogue dynamics.
We apply 
four existing linguistic annotations to subsets of SCOUT to advance understanding of phenomena within human-robot dialogue. To this end, we annotate SCOUT with (1) Abstract Meaning Representation (AMR) \cite{banarescu2013abstract} and (2) Dialogue-AMR \cite{bonial2019abstract} to assess interlocutor intents and meaning within an utterance. To track relationships between utterances, we annotate both (3) Transactional Units (TUs) \cite{carletta-etal:1996} and (4) the Relations between utterances within a TU; together these reveal patterns of the Dialogue Structure \cite{traum2018dialogue}.
We describe how these annotations on top of SCOUT have been subsequently used in action selection for automated robot navigation systems 
and automated dialogue systems 
along with other areas of human-robot analysis.

This paper offers the following contributions:
\begin{itemize}
        \vspace{-0.07in}
  \item A fully-compiled, novel corpus of multi-modal, robot-directed dialogue in a collaborative exploration task involving a remotely-located physical or simulated robot (Sec.~\ref{sec:corpus-summary} and~\ref{sec:corpus-stats}).
  \item Annotations of SCOUT data using existing frameworks of AMR, Dialogue-AMR, and Dialogue Structure TUs and Relations (Sec.~\ref{sec:annotations}).
  \item Applications of SCOUT data and annotations, including the development of systems and analyses of the language and behaviors observed in the human-robot dialogue (Sec.~\ref{sec:applications}).
\end{itemize}

Although portions of the dialogues have previously been shared via private data-sharing agreements to enable the annotations and applications described here, this paper presents the comprehensive compilation and curation process of the dialogues and multi-modal streams for public release
under a Creative Commons Zero 1.0 Universal (CC0 1.0) license. We believe the corpus and its annotation will serve to immediately benefit the Human-Robot Interaction (HRI), Dialogue, and broader Robotics communities that aspire to use language as a way to interact with robots. SCOUT is available at \url{https://github.com/USArmyResearchLab/ARL-SCOUT}

\section{Experiments and Data Collection}

\begin{table*}[ht]
\small
\null\hfill 
\begin{tabular}{c|c|c|c|c} \toprule
 & {\bf Experiment 1} & {\bf Experiment 2} & {\bf Experiment 3} & {\bf Experiment 4} \\
\midrule
Dialogue Processing & WoZ + keyboard & WoZ + button GUI & WoZ + button GUI & ASR + auto-DM\\
Robot Behaviors & WoZ + joystick & WoZ + joystick & WoZ + joystick & WoZ + joystick\\
Robot \& Environment & physical & physical & virtual & virtual
\\ \bottomrule
\end{tabular}
\hfill\null 
\caption{Human-Robot Dialogue Experimentation. ASR:\ Automatic Speech Recognition}
\label{expt-phases}
\end{table*}

The experimental domain involved a collaborative human-robot exploration task in a low-bandwidth environment \cite{marge2016applying}. The robot (a Clearpath Jackal with functionality implemented in ROS, the Robot Operating System \cite{koubaa2017robot}) entered an unexplored area and received instructions from a remotely-located human teammate. This human teammate (called "Commander") was given specific goals for the exploration, such as locating and counting doors and specific objects of interest, e.g., doorways, shovels, shoes. The Commander could not directly teleoperate the robot and had to provide verbal instructions to accomplish tasks with the robot (e.g., ``Move forward five feet'', ``Proceed through the doorway in front of you''). The Commander's knowledge of the environment was based upon (1) a dynamic LIDAR map of the area built up in real time as the robot moved, (2) snapshot pictures from the robot's front-facing RGB camera, taken upon Commander request, and (3) text messages from the robot. The left-hand side of Fig.~\ref{fig:method} depicts what the Commander saw and could do during the interaction while seated at their workstation.

The robot was controlled using a Wizard of Oz methodology to facilitate a data-driven understanding of how people talk to robots~\citep{riek2012wizard,devault2014simsensei}.
The experiments employed two Wizards; their division of labor is depicted on the right-hand side of Fig.~\ref{fig:method}. The Dialogue Manager Wizard (DM-Wizard) listened to the Commander's instructions and decided how the robot should proceed in the dialogue. Status updates and clarifications were sent to the Commander from the DM-Wizard in a chat window. When the DM-Wizard determined that instructions were executable in the current context, the instructions were passed in a constrained form 
to the Robot Navigator Wizard (RN-Wizard), who used a joystick to teleoperate the robot and who provided information of failures through speech or a button click back to the DM-Wizard as the task was being completed. The DM-Wizard passed status updates from the RN-Wizard to the Commander.

Four experiments varied the modes of dialogue processing 
and how the robot and environment in the experiments were represented (Table~\ref{expt-phases}).
\textbf{Experiment~1} had a DM-Wizard manually type messages to interact with the Commander in real-time, and a RN-Wizard control a physical robot in environments that resembled an alleyway or an indoor space of a house-like
environment under construction. The house contained a variety of hallways, rooms, and
objects that gave themes to different spaces, for example, a kitchen area, conference room, and office.

\textbf{Experiment 2} automated the DM-Wizard's command handling and response generation with a click-button graphical user interface (GUI). The collection of messages uncovered from Experiment 1 was incorporated into the GUI for use by the DM-Wizard, substantially reducing typing and composition effort by the DM-Wizard while increasing response uniformity. 
The GUI design built in functionality to avoid inflexible situations by using open slots where the DM-Wizard could type in a value for well-defined templates, e.g., “I see a door on the left,” “I see a door on the right,” “I see a wall,” etc. There were no entirely open response buttons; all buttons reflect, at a minimum, an observed template of responses like “I see \_\_\_". \cite{bonial2017laying}. Experiment 2 took place in the same physical environments as Experiment 1. 

\textbf{Experiment 3} utilized the same DM-Wizard GUI and moved from a physical robot and environment, to a simulated one designed to be a 1--1 replica of the physical environments, including the objects and their placement. Gazebo, a high fidelity 3D simulator, was used to construct the environment and complete the experiments~\citep{koenig2004design}. The simulated robot was programmed with the same capabilities as the physical one using ROS. From the Commander's perspective, the study was equivalent to the previous two, with the exception that the images from the camera were virtually rendered.

Finally, \textbf{Experiment 4} deployed a completely automated dialogue system trained on data collected from the prior experiments.  
Instead of a DM-Wizard listening to the Commander's speech and routing
messages back and to the RN-Wizard, the dialogue
system provided these capabilities. This auto-DM system was divided into the following components: (1) an automated speech recognition (ASR) component that would 
transcribe speech in real time from the Commander, (2) a dialogue manager including a classifier that would determine the Commander's intent from their speech using training data collected in the previous experiments, and determine whether to (3) translate instructions to the RN-Wizard, and/or (4) provide replies to or request clarification from the Commander. This experiment had only the RN-Wizard as a wizard experimenter to tele-operate the robot in response
to instructions provided to it by the auto-DM system. 
The auto-DM system was trained on Experiment 1 and 2 data, and tested on a subset of Experiment 3 data. Several training methods were employed, and accuracy ranged from 61\% - 75\%. Over half of the incorrect responses would still be appropriate and advance the dialogue, as they were considered felicitous (appropriate responses that would have the same effect as the correct response) or approximate (responses that differed only slightly from the correct one, e.g., variation in turn radius or movement distance) \cite{gervits2019classification}.

\section{SCOUT Construction}
\label{sec:corpus-summary}

In this section, we describe our methodology to isolate, extract, and verify the various data streams from the experimentation, and our process to compile the information into the human-readable and machine-processable formats that comprise SCOUT.

\textit{Commander data} consists of the Commanders' verbal interactions with the robot. These were recorded in Mumble\footnote{\url{https://www.mumble.info/}} using a push-to-talk button. Annotators were trained to manually transcribe the Commander speech from Experiments 1--3 using the Praat software \cite{praat}. Turns were segmented by silences, and then further by intents. For example, the turn ``go to the map and take a picture'' would be segmented into the utterances ``go to the map'' and ``and take a picture'' \cite{bonial2019transcription}.
The speech data from Experiment 4 was transcribed during the experiment by Google ASR or Kaldi. Key tokens were automatically normalized, for example, converting ``five ft'' to ``5 feet'' for consistency with the manual transcription coding schema in Experiments 1--3.
Post-experimentation, manual verification was conducted to correct ASR results, for example ``turn right for you to grieve'' was corrected to ``turn right 45 degrees'' after listening to the utterance. The timestamp at which the utterance began (which was not necessarily the same as when the push-to-talk button was depressed) was saved in metadata by Mumble or the ASR for future alignment.

\begin{table*}[ht]
\small
\begin{center}
\begin{tabular}{llp{1.4in}p{1.1in}p{1.0in}p{0.8in}}
\toprule
   & & \multicolumn{2}{c}{Left Conversational Floor}                               & \multicolumn{2}{c}{Right Conversational Floor}           \\
   \cmidrule(r{1em}l{0.75em}){3-4}
   \cmidrule(rl{0.75em}){5-6}
\textbf{\#ID} & \textbf{Timestamp} & \textbf{CMD}      & \textbf{DM $\rightarrow$ CMD} & \textbf{DM $\rightarrow$ RN} & \textbf{RN}   \\
\midrule
222 & 1054.31 & robot proceed through the doorway &  &  & \\
223 & 1061.9 & turn a hundred and eighty degrees to the right &  &  & \\
224 & 1063.78 & and take a picture &  &  & \\ \midrule
225 & 1070.54 &  & processing. . . &  & \\
226 & 1077.99 &  & I see more than one doorway. &  & \\
227 & 1079.46 &  & The one to my left? &  & \\ \midrule
228 & 1081.34 & the doorway to your left &  &  & \\ \midrule
229 & 1085.72 &  & processing. . . &  & \\ \midrule
230 & 1103.95 &  &  & move to Foyer - Kitchen doorway & \\
231 & 1107.81 &  &  & then. . . & \\
232 & 1109.87 &  &  & turn 180 & \\
233 & 1111.06 &  &  & then. . . & \\
234 & 1112.19 &  &  & send image & \\ \midrule
235 & 1114.76 &  & moving. . . &  & \\
236 & 1121.65 &  & turning. . . &  & \\ \midrule
237 & 1134.48 &  &  &  & done and sent \\ \midrule
238 & 1135.29 &  & done, sent &  & \\
\bottomrule
\end{tabular}
\end{center}
\caption{Navigation instruction initiated by the Commander (\#222-244), clarification (\#225-229), translation to a simplified form by Dialogue Manager (DM) to Robot Navigator (RN) (\#230-234), status updates (\#235-236), completion by the RN (\#237), and notification of task completion to Commander (\#238).}
\label{tab:tu}
\end{table*}

\textit{Dialogue Manager data} are text messages sent by the DM to either the Commander or the RN-Wizard. Text messages were recorded as a modified \texttt{sensor\_msgs/StringStamped} ROS topic in a ROS bag file---a file format also used to save diverse timestamped sensor data from the robot. Text messages to the Commander and to the RN-Wizard were differentiated from each other in the ROS bag file. The timestamps were extracted along with the utterances for  alignment.

\textit{Robot Navigator data} consists of the RN-Wizard's verbal or text messages to the DM. In Experiments 1--3, RN communications were spoken and transcribed following the same process using Praat, saving the timestamps from the metadata. In Experiment 4, communications were text messages the RN triggered by a button press on the joystick coded as the \texttt{sensor\_msgs/Joy} ROS topic, or on the GUI coded as the \texttt{sensor\_msgs/StringStamped} ROS topic. Text messages and their timestamps were extracted from the ROS bag file for alignment.

The transcribed speech and text utterances from all interlocutors were time-aligned into communication floors that reflect the passing of information during the dialogue as it occurred. The streams are shown in Table~\ref{tab:tu}. Given that the DM served as an intermediary directing communications between the Commander and RN, the dialogue took place across two non-mutual conversational floors: the Left and Right floors.
The Left floor includes communication between the Commander and what the Commander thinks of as ``the robot,'' (really the DM, acting as front end), and contains streams ``CMD'' and ``DM$\rightarrow$CMD.'' The Right floor was between the DM and RN, and contains streams ``DM$\rightarrow$RN'' and ``RN.'' The DM would convey information across floors, as shown in lines 230-234 (Left to Right) and 238 (Right to Left). Timestamps were coded either as seconds since the dialogue started (Experiments 1 and 2) or with a unix timestamp (Experiments 3 and 4). Complications arose in 
synchronizing the timestamps across recordings from the three different machines used to run the experiments (one each for Commander, DM, and RN), and required scripts and manual verification to ensure that each utterance was inserted into the correct location in the transcript across the conversational floors. 

The resultant \textit{time-aligned transcripts\/} were compiled into .xlsx spreadsheets in the format of Table~\ref{tab:tu} (see Fig.~\ref{fig:transcript} in Appendix for a screenshot),
as well as a compacted tab delimited format to facilitate different methods of file-processing: 
\begin{footnotesize}
\begin{verbatim}
ID  time    stream   text     
...
222	1054.31 CMD     "robot proceed 
                    through the doorway"
223	1061.9  CMD     "turn a hundred and 
                    eighty degrees to 
                    the right"
224	1063.78 CMD     "and take a picture"
225	1070.54 DM->CMD "processing. . ."
...
\end{verbatim}
\end{footnotesize}
In Experiment 4, the .xlsx spreadsheets contain additional columns for the raw ASR results and intermediary normalized forms, in addition to the final corrected utterance (see Fig.~\ref{fig:transcript4} in Appendix). The Commander text in these tab delimited formats is the corrected utterance.

\textit{Images\/} were taken by the RN and recorded as \texttt{sensor\_msgs/Image} ROS topics in ROS bag files. After extraction, each image per dialogue was given a unique id and inserted into a modified .xlsx spreadsheet at the moment the RN-Wizard had the robot take the image (see Fig.~\ref{fig:iscout} in Appendix.) 
In the spreadsheet, clicking or control-clicking on the image name will open the .jpg image in the computer's default picture viewer program. This information is also available in a modified tab delimited format, where the stream value is ``IMAGE'' and the text value is the image filename: 

\begin{footnotesize}
\begin{verbatim}
ID    time    stream  text              
...
234	 1112.19  DM->RN  "send image"	
235	 1114.76  DM->CMD "moving. . ."		
236	 1121.65  DM->CMD "turning. . ."
i019 1133.96  IMAGE	  "frame019"
237	 1134.48  DM->RN  "done and sent"
...
\end{verbatim}
\end{footnotesize}

\begin{figure*}[ht!]
    \centering
    \begin{subfigure}[t]{0.34\textwidth}
        \includegraphics[width=\textwidth]{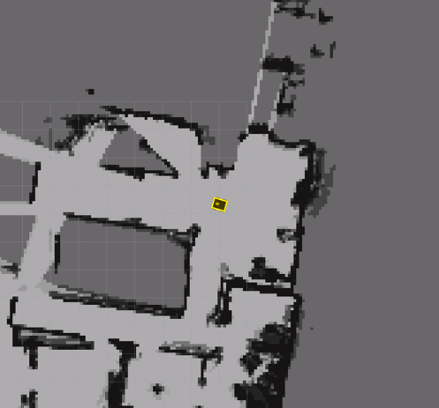}
        \caption{\label{fig:endmap}LIDAR map at the end of a dialogue. Dark gray is unscanned by LIDAR}
    \end{subfigure}
    \begin{subfigure}[t]{0.24\textwidth}
    \centering
        \includegraphics[width=0.8\textwidth]{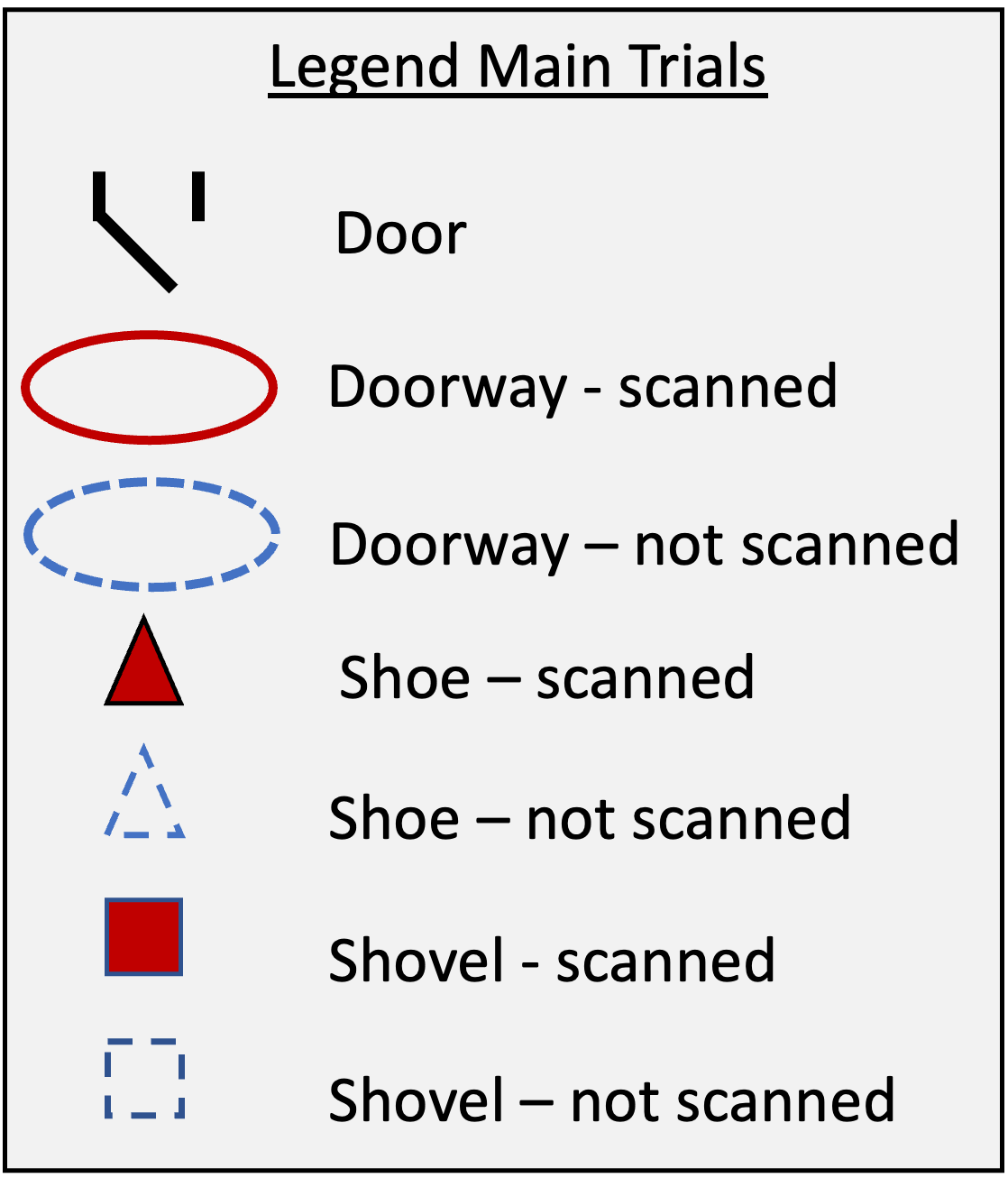}
        \caption{\label{fig:legend}Legend for Fig.~\ref{fig:floorplan}}
    \end{subfigure}
    \begin{subfigure}[t]{0.37\textwidth}
        \includegraphics[width=\textwidth]{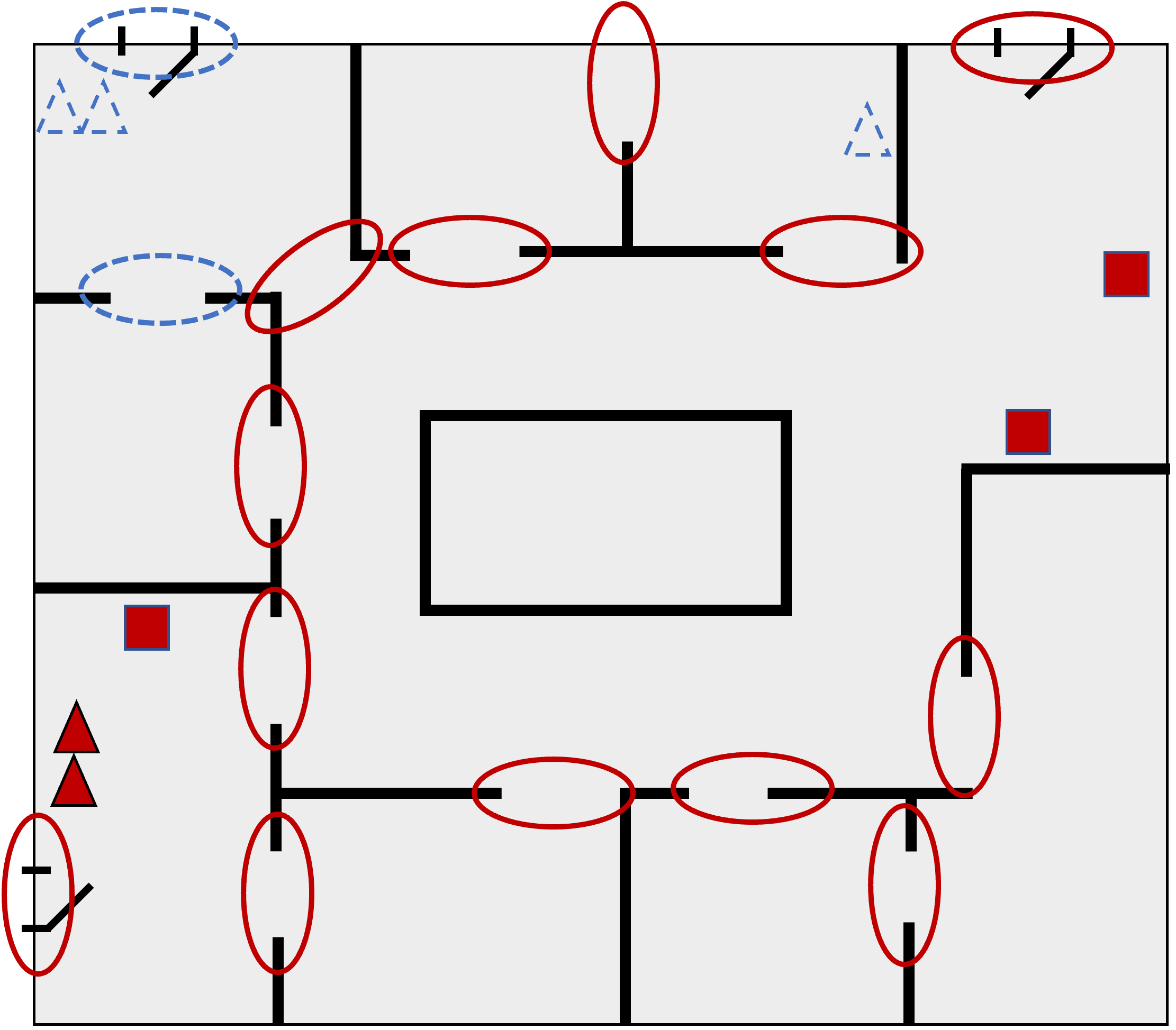}
        \caption{\label{fig:floorplan}Corresponding top-down floor plan of Fig~\ref{fig:endmap} showing items' scanned status}
    \end{subfigure}
            \vspace{-0.1in}
    \caption{One LIDAR map and annotated floor plan with items scanned or not scanned by the LIDAR marked. Floor plan and legend were not shown during the exercise.}
    \label{end-map}
\end{figure*}

\textit{LIDAR maps\/} were extracted from what the LIDAR had scanned by the end of each dialogue (Fig.~\ref{fig:endmap}). These are snapshots from the last frame of the dialogue as captured in the screen recording of the Commander's monitor.
Due to how the LIDAR visualization functionality was implemented, some snapshots show a subset of the environment rather than a complete view. One .png file was extracted for each dialogue in Experiment 1, with future plans to streamline the process and compile maps for the remaining dialogues. 

\textit{Rendered floor plans\/} of the complete environment were manually created for each of the dialogues with LIDAR maps, showing the objects of interest scanned by the LIDAR (Fig.~\ref{fig:floorplan}, with legend in Fig.~\ref{fig:legend}). These determinations were made by manually comparing the LIDAR to the rendered floor plan. Dark gray spaces indicated the LIDAR had not scanned the area and informed the determinations. A text version of this annotated floor plan is also available, where each target entity (e.g., doorway, shovel, etc.) was mapped to a unique identifier (e.g., ``door1'') and denoted as scanned or not scanned:
\begin{footnotesize}
\begin{verbatim}
door1    not-scanned
door2    scanned
...
\end{verbatim}
\end{footnotesize} 
            \vspace{-0.2in}

\section{SCOUT Statistics}
\label{sec:corpus-stats}

The corpus contains data from 93 Commanders, where each Commander completed three dialogues---one training exercise in the alleyway and two main exercises in the house, starting at different locations with different objects of interest to count---for a total of 278 dialogues containing, on average, 320 utterances.\footnote{One Commander only completed the training and  one main, thus SCOUT has 278 and not 279 dialogues.} The corpus contains 89,056 utterances total, 310,095 words, and 5,785 images (Table~\ref{tab:statistics-main}). On average, Commanders requested 20.8 images per dialogue which they could use to assess the environment in their search and counting tasks. On a per individual and per task basis (training or main), the requests ranged from 3--88 images. These statistics per experiment are given in Table~\ref{tab:statistics-extra} in the Appendix.

\begin{table}[]
    \centering
    \begin{tabular}{rl}
        \toprule
         \textbf{Corpus Attribute} & \textbf{SCOUT Total}  \\ \midrule
         Commanders & 93 \\
         Dialogues & 278 \\
         Utterances & 89,056 \\
         Words -- All & 310,095\\
         Words -- Unique & 89,056 \\
         Images & 5,785 \\ 
         Avg. Utterances per Dialogue & 320 \\
         Avg. Images per Dialogue & 20.8 \\ \midrule
         Standard-AMR Sentences & 569 \\
         Dialogue-AMR Sentences & 569 \\ 
         Dialogue Structure TUs & 13,663 \\
         Dialogue Structure Relations & 69,430 \\
         \bottomrule
         
    \end{tabular}
    \caption{SCOUT corpus summary. Corpus attributes per experiment in Table~\ref{tab:statistics-extra} in Appendix.}
    \label{tab:statistics-main}
\end{table}

Table~\ref{tab:statistics} shows the breakdown of corpus attributes by experiment and interlocutor per conversational floor. We also tabulated a subset of 30 randomly selected dialogues from Experiment~3 in order to show a more fair comparison across experiments due to the difference in participant pool size (shown in parenthesis next to the full counts in the Experiment~3 column.) 

The Commanders spoke a total of 25,386 utterances (2,446 unique words) 
and the RN-Wizards spoke 13,805 utterances (526 unique words) as shown in ``Total'' column in Table~\ref{tab:statistics}. 
The difference in counts is likely due to their roles in the experiment. The primary vocabulary of the RN-Wizard comes from acknowledgements of DM-wizard requests and reports of failures, whereas the Commander vocabulary reflects their attempts to take initiative and issue the requests. As a result, we observe the Commander vocabulary is greater and more varied than the RN-Wizard.

The Dialogue Manager sent a total of 31,959 text messages (813 unique words) to the Commander (DM$\rightarrow$CMD), and sent 17,906 text messages (537 unique words) to the RN (DM$\rightarrow$RN). We observe that the variety in vocabulary drops from Experiment~1 to Experiment~2, which likely reflects the introduction of the GUI (``DM$\rightarrow$CMD Words -- Unique`` 565 to 311; and ``DM$\rightarrow$RN Words -- Unique`` 349 to 141 in Table~\ref{tab:statistics}). 
Aspects of the DM-Wizard's dialogue processing were compared in Experiments~1 and~2 (i.e., keyboard vs. button GUI) for assessing the Commander's ability to work with the DM-Wizard to issue well-formed and executable instructions. In combination with the dialogue structure analysis in Section~\ref{sec:annotations}, we found that, compared to Experiment~1, more instructions were issued in Experiment~2 within the same 20 minute trial limit, and more instructions were successful, showing improvement in the speed of the interaction \cite{marge2018balancing}. The corpus attributes remain reasonably consistent within Experiment~1 and~2 values for the Experiment~3 Subset.

In Experiment~4, we observe a significant increase in DM$\rightarrow$CMD frequency of words from the prior experiments (``DM$\rightarrow$CMD Words -- All'' 5,550--10,923 in Exps~1--3Subset, up to 24,253 in Exp~4), perhaps due to the introduction of the auto-DM, while the Commander variety of words decreases (``CMD Words -- Unique`` 661--738 in Exps~1--3Subset, down to 320 in Exp~4). We suspect the rise in frequency signifies an increase in miscommunication with the auto-DM, and that the lack of Commander vocabulary variety, while still maintaining the same level of word frequency, is due to an attempt to revert to more simplified terms, or hesitation to `try out' different ways of giving instructions due to the auto-DM's limitations; the word error rate in Experiment 4 was 25\%.

\begin{table*}[ht!]
\begin{footnotesize}
	\centering
	\begin{tabular}{rlrrrrrr}
	\toprule
		 \multicolumn{2}{c}{{\bf Corpus Attribute}} & {\bf Exp. 1} & {\bf Exp. 2} & {\bf Exp. 3 (Subset)} & {\bf Exp. 4} & {\bf Total} \\
            \midrule
            & Dialogues & 30 & 30 & 188 (30) & 30 & 278 \\ \midrule
            CMD &
            Utterances & 1,819 & 2,161 & 18,206 (2,545) & 3,200 & 25,386 \\
            & Words -- All & 9,883 & 10,923 & 85,549 (11,453) &10,633 & 116,988 \\
            & Words -- Unique & 738 & 661 & 2,078 (675) & 320 & 2,446 \\ 
            & Avg. Words per Utterance & 5.43 & 5.05 & 4.70 (4.50) & 3.32 & 4.61 \\ 
            \midrule

            DM$\rightarrow$CMD & Utterances & 1,779 & 3,370 & 20,595 (2,987) & 6,215 & 31,959 \\ 
            & Words -- All & 5,550 & 10,923 & 65,889 (10,485) & 24,253 & 106,615 \\
            & Words -- Unique & 565 & 311 & 418 (326) & 335 & 813 \\ 
            & Avg. Words per Utterance & 3.12 & 3.24 & 3.20 (3.51) & 3.90 & 3.34 \\ \midrule
            
            DM$\rightarrow$RN &  
            Utterances & 1,417 & 1,766 & 12,622 (1,688) & 2,061 & 17,906 \\ 
            & Words -- All & 5,139 & 5,588 & 41,038 (5,568) & 7,433 & 59,196 \\
            & Words -- Unique & 349 & 141 & 289 (158) & 51 & 537 \\ 
            & Avg. Words per Utterance & 3.63 & 3.16 & 3.25 (3.30) & 3.61 & 3.31 \\ \midrule

            RN & Utterance & 1,082 & 1,124 & 8,436 (1,042) & 3,163 & 13,805  \\
            & Words -- All & 2,246 & 1,647 & 14,528 (1,935) & 8,875 & 27,296 \\
            & Words -- Unique & 253 & 39 & 349 (93) & 105 & 526 \\ 
            & Avg. Words per Utterance & 2.08 & 1.47 & 1.72 (1.86) & 2.81 & 1.98 \\ 

        \bottomrule
	\end{tabular}
	\caption{SCOUT corpus statistics by experiment and interlocutor and conversational floor \label{tab:statistics}}
\end{footnotesize}
\end{table*}

\section{Annotations on SCOUT}
\label{sec:annotations}

With SCOUT fully assembled, we applied existing linguistic annotation schemas in order to better understand how humans worked with the robot, namely through analyzing the form and content of their instructions. We include in SCOUT's release Abstract Meaning Representation (AMR), Dialogue-AMR, and Dialogue Structure TU and Relation annotations (quantities shown in Table~\ref{tab:statistics-main}). Taken together, these make different levels of conversational patterns accessible to automated systems---propositional semantics of an utterance (AMR), the illocutionary force (Dialogue-AMR), the meso-level intentional structure of a set of utterances (Dialogue Structure TUs), and finally the individual relations of each subsequent utterance within a TU to an antecedent utterance (Dialogue Structure Relations)\footnote{See references for details of annotation schemas.}.

\subsection{Standard-AMR and Dialogue-AMR Annotation}

To distill a robot's behavior primitives and their parameters from totally unconstrained natural language, we 
apply AMR, a formalism for sentence semantics that abstracts away many syntactic idiosyncrasies and represents sentences with directed, acyclic graphs \cite{banarescu2013abstract}. Below is the utterance ``take a photo'' represented in Standard-AMR form in PENMAN representation \cite{penman1989}; note that AMR abstracts away from the semantically light verb ``take'', instead representing photographing semantics:
\begin{footnotesize}
\begin{verbatim}
(p / photograph-01
        :ARG0 (y / you)
        :mode imperative)
\end{verbatim}
\end{footnotesize}

Dialogue-AMR is an enhanced AMR representing not only the content of an utterance, but the illocutionary force behind it, as well as tense, aspect, and completability \cite{bonial2019abstract}---all aspects of meaning that are necessary for the robot to interpret and act upon the natural language instructions. The same utterance as above is represented in Dialogue-AMR form using a domain-specific action frame \texttt{send-image-99} which represents a robot's photographing behavior:
\begin{footnotesize}
\begin{verbatim}
(c / command-SA
        :ARG0 (c2 / commander)
        :ARG1 (s / send-image-99
            :ARG0 (r / robot)
            :ARG1 (i / in-front-of
                :op1 r)
            :ARG2 c2
            :completable +
            :time (a / after
                :op1 (n / now)))
        :ARG2 r)
\end{verbatim}
\end{footnotesize}
We annotated subsets of utterances from Experiments 1 and 2 with Standard-AMR and Dialogue-AMR, and an entire dialogue from Experiment 3. These annotations are available in parallel .txt files where each annotated utterance is given a unique id corresponding to the dialogue and sentence ID from the .xlsx corpus files. 

\subsection{Dialogue Structure Annotation}

\begin{figure*}[!ht]
\begin{footnotesize}
\begin{verbatim}
ID  time    stream   text                                 TU  ant.  relation
...
222	1054.31 CMD      "robot proceed through the doorway"  28  None  None	
223	1061.9  CMD      "turn a hundred and eighty degrees   28  222   continue
                     to the right"                   
224	1063.78 CMD      "and take a picture"                 28  223   continue
225	1070.54 DM->CMD  "processing. . ."                    28  224*  processing
226	1077.99 DM->CMD  "I see more than one doorway."       28  222   missing-info
227	1079.46 DM->CMD  "The one to my left?"                28  222   req-clar
228	1081.34 CMD      "the doorway to your left"           28  227   clar-repair
...
\end{verbatim}
\end{footnotesize}
\vspace{-0.1in}
\caption{Tab delimited format for Dialogue Structure. Transaction Unit (TU), antecedent (ant).}
\label{fig:format-ds}
\vspace{-0.1in}
\end{figure*}

To understand and make explicit the patterns of multi-floor conversation, we applied Dialogue Structure annotation to capture the relationships between utterances within and across the conversational floors. Each aligned transcript was divided into Transaction Units (TUs) defined in \citet{traum2018dialogue} as clusters of utterances across conversational floors that sequentially work towards fulfilling the original speaker's intent. A TU may encompass multiple dialogue utterances, spanning multiple speaker turns, including requests for clarification and subsequent repairs, confirmations and various types of acknowledgments that instructions were heard, understood, and complied with.  Each utterance was further annotated with the {\it Relation} and {\it Antecedent}---the ID of the most immediate direct relation between this utterance and a prior utterance \cite{traum2018dialogue}. Any contextual information required for understanding the annotation was denoted. 

Every dialogue in SCOUT was annotated with this formalism and recorded in new .xlsx spreadsheets (Fig.~\ref{fig:annotations} in Appendix) 
and converted into the tab delimited format in Fig.~\ref{fig:format-ds}. In this example, \#222 is the start of a new TU and assigned no relation. The instruction is continued by the same interlocutor into \#223 and \#224 through the {\it continue} relation. In \#226 the DM informs the Commander that information is missing ({\it missing-info}) for successful execution of the instruction, and thus their request for a clarification ({\it req-clar}), to which the Commander provides the appropriate repair ({\it clar-repair}) in \#228.

\section{Applications}
\label{sec:applications}

SCOUT and its annotations provide for a variety of analyses in support of Robotics research, especially within the HRI and Dialogue communities. We briefly describe research directions making use of SCOUT for system development, and how the data have encouraged discovery of new questions on how humans speak to robots. 

\subsection{Systems Developed from SCOUT Annotations}

The SCOUT annotations have been used to train and deploy fully autonomous dialogue and navigation prototypes. The \textit{ScoutBot} system utilizes the Dialogue Structure annotations from Experiments 1--3 to enhance the auto-DM developed in Experiment 4, and further implement autonomous robot navigation through ROS twist messages that map to user intents in a simulated building \cite{lukin2018pipeline,gervits2019classification}. The \textit{MultiBot} system extends this auto-DM pipeline to a simulated urban outdoor environment with multiple robots, further integrating heuristics for goal-based navigation instructions \cite{marge-etal-2019-research}. The AMR and Dialogue-AMR annotations have been utilized for designing a classifier for intents, and integrating with a Clearpath Husky Unmanned Ground Vehicle in the real-world in robot-directed navigation \cite{bonial-dmr-2023}. The SCOUT corpus was also used to train a dialogue structure parser \cite{kawano2023end}.

\subsection{Analyses of Human-Robot Multi-Modal Communication}

The diversity of Commanders and the open-ended nature of the communication gives rise to many questions about Commander instruction-giving and navigation preferences. Researchers have asked how Commanders' interactions with respect to time and trust affect their instruction-style, and found an increase in landmark instructions (e.g., ``move to the door in front of you'') over metric information (e.g., ``move forward five feet'') \cite{marge2017exploring} as well as more verbose and compound instructions over time and with increasing trust \cite{lukin2018consequences}. \citet{moolchandani2018evaluating} used the navigation patterns observed in SCOUT to discover that humans prefer when the robot demonstrates a sense of self-safety and awareness of its environment.
The multi-modal nature of the exchanges has been explored, finding a relationship between the success of item-counting and exploration and quantity of images taken \cite{lukin2023navigating}.
The corpus has also allowed for study of different types of capabilities robots should have to conduct natural dialogue-based interactions with humans \cite{pollard2018we}, and the exchanges between the Commander and Dialogue Manager have been evaluated for various linguistic modalities (i.e., presence of modal expressions, negation, and quantifiers) \cite{donatelli-etal-2020-two}.

The images of SCOUT represent an opportunity to develop new computer vision and language techniques. A subset of these low-resolution and dim images with atypical angles have been used in computational visual storytelling to represent diverse environments and presentation of imagery 
found to be lacking in other collections of visual storytelling data \cite{lukin2018pipeline}. 
In analyzing a human-authored story collection utilizing SCOUT images, \citet{halperin2023envisioning} observed narrative biases with respect to the cultural and linguistic biases associated with what human-authors recognized in the images.

\section{Related Work}

There are many human-human situated corpora that exhibit the Director-Follower paradigm SCOUT follows. 
These corpora have been used to study referring expressions \cite{stoia2008scare,liu2016coordinating,hu2016entrainment}, speaker intents \cite{narayan2019collaborative,bonial2021builder}, structure \cite{eberhard2010indiana}, and to develop autonomous robot systems for following directions \cite{macmahon2006walk,chen2011learning,de2018talk,suhr2019executing,chen2019touchdown,thomason2020vision,padmakumar2021teach,gervits2021should}.
Other situated paradigms leverage knowledge about object affordances within the world, e.g., a cup observed to be on its side may roll, and in combination with human gestures, inform a robot's reasoning and actions within simulated spaces \cite{pustejovsky2020situated}.
However, prior work has shown that the way humans instruct robots is different from how they instruct other humans 
\citep{mavridis2015review,marge2020let,MARGE2022101255}. There is a critical need for corpora like SCOUT that capture the human-robot dynamic in a coordinated Director-Follower task.

The Multi-Woz corpus \cite{budzianowski_mihail_rahul_shachi_sethi_agarwal_gao_hakkani-tur_2019} is the largest of several corpora (including \citet{eric2017key}) collected in a WoZ crowd-sourcing paradigm  proposed in \citet{wen2016network}. Here, each crowd-worker acted as either a wizard or a user and supplied only a single turn after observing previous turns. While this paradigm allows for scaling to a larger training corpus, it is unclear that this turn-by-turn addition to the dialogue in text via an online portal can reveal naturalistic dialogue patterns or individual communication style differences.
Therefore SCOUT presents a unique resource for studying multi-modal and situated dialogue within a more natural interaction modeled between a human and robot.

Recent zero-shot approaches using large language models show the ability to process robot-directed instructions and generate an executable plan without needing a corpus or annotations (e.g., \citet{brohan2023can}). Yet because these models are not trained on domain experience, they cannot afford the same rich semantic and structural knowledge supplied by SCOUT data and annotations.

\section{Conclusion}

SCOUT meets the characteristics outlined at the start of this paper for datasets to study human-robot dialogue. The corpus focuses on how humans would instruct a robot (rather than another human) to perform navigation tasks, and how robots could respond in a variety of situations. It explores the natural diversity of communication strategies in situated dialogue, ranging from complex, abstract-level instructions to lower-level basic control.
The data and annotations have been used to advance our understanding of human-robot dialogue and to develop automated robotic systems. 

We envision this corpus as providing critical annotation infrastructure and insight into multi-party cooperative tasks, for example, between heterogeneous human-robot teams. Instead of the three interlocutors (CMD, DM, RN) speaking across two conversational floors in SCOUT, a heterogeneous team of robots could organize communication with human participants in different ways, for example, unique conversational floors between CMD and each robot to avoid channel contention, or the robots communicating to each other within a conversational floor from which to then report back to the CMD. 
The corpus and its annotations thus represent one possible configuration out of many for multi-modal, multi-party human-robot interaction for studying how visual information, intents, and goal progress is tracked.

\section{Acknowledgements}

We thank many individuals for their past contributions to the project: Carlos Sanchez Amaro, David Baran, Jill Boberg, Brendan Byrne, Austin Blodgett, Taylor Cassidy, Jessica Ervin, Arthur William Evans III, Carla Gordon, Jason Gregory, Darius Jefferson II, Brecken Keller, Su Lei, Elia Martin, Pooja Moolchandani, John J Morgan, Reginald Hobbs, Brandon Perelman, John G Rogers III, Douglas Summers-Stay, Adam Wiemerslage, Volodymyr Yanov. We also thank our reviewers for suggestions on how to strengthen this paper.

\section*{Ethical Considerations}

This data was collected following an approved IRB protocol with all participants signing a consent form. Personally identifiable information (PII) that was recorded during the study (i.e., participant speech and recordings of their face) are not released in SCOUT. Additional permissions to use PII, including presenting audio/video clips at conferences or publicly releasing the full audio/video, was agreed to with explicit permission.

\section*{Limitations}

Due to the choice of experimental design, SCOUT data may not generalize to all scenarios. For instance, the vocabulary may be only representative of the search task assigned to the Commanders, and the low-lighting in the images may prove challenging for state-of-the-art computer vision algorithms trained on `canonical' environments.

\section{Bibliographical References}\label{sec:reference}

\vspace{-0.2in}

\bibliographystyle{lrec-coling2024-natbib}
\bibliography{LREC-BIB}

\onecolumn  
\appendix
\section*{Appendix}
\begin{figure*}[h]
    \centering
    \includegraphics[width=\textwidth]{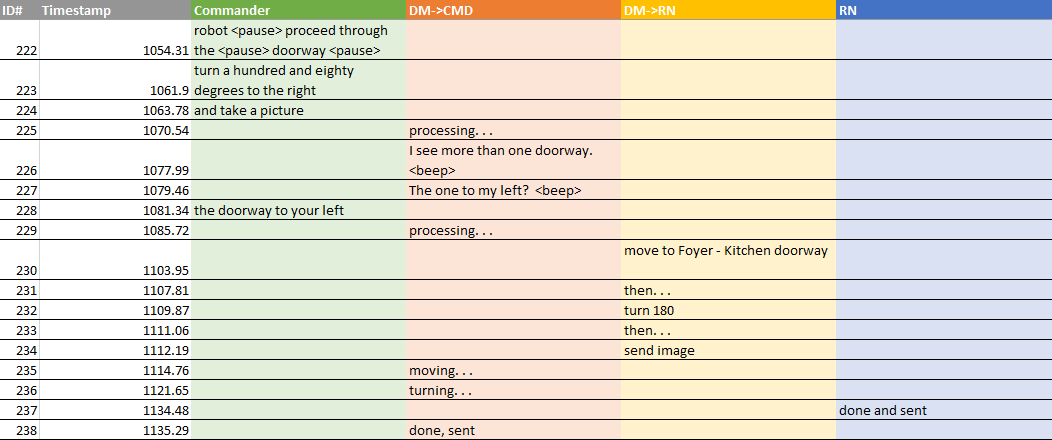}
    \caption{Aligned .xlsx transcript screenshot format for Experiments 1-3}
    \label{fig:transcript}
\end{figure*}

\begin{figure*}[t]
    \centering
    \includegraphics[width=\textwidth]{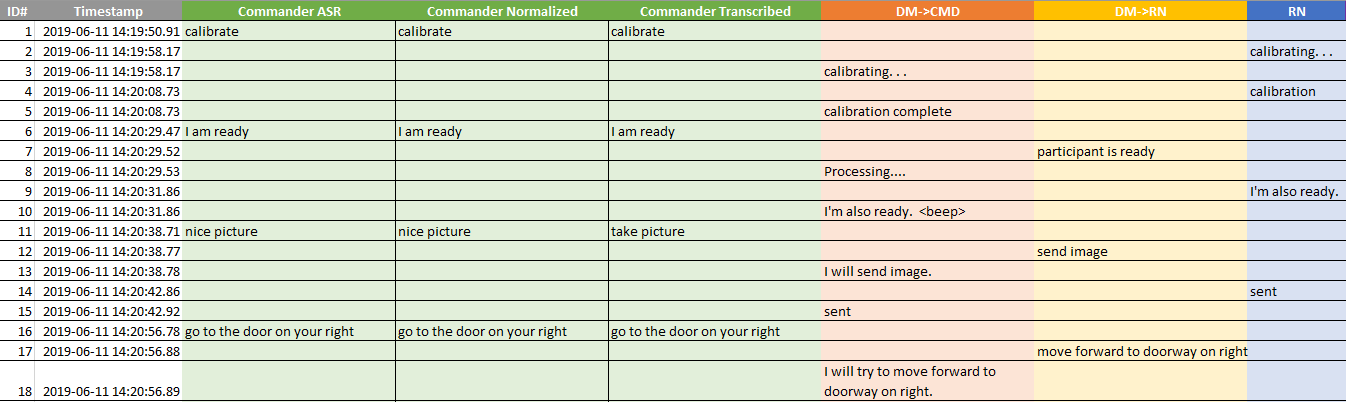}
    \caption{Aligned .xlsx transcript screenshot format for Experiment 4 with the ASR results and intermediary normalized forms}
    \label{fig:transcript4}
\end{figure*}

\begin{figure*}[t!]
    \centering
    \includegraphics[width=\textwidth]{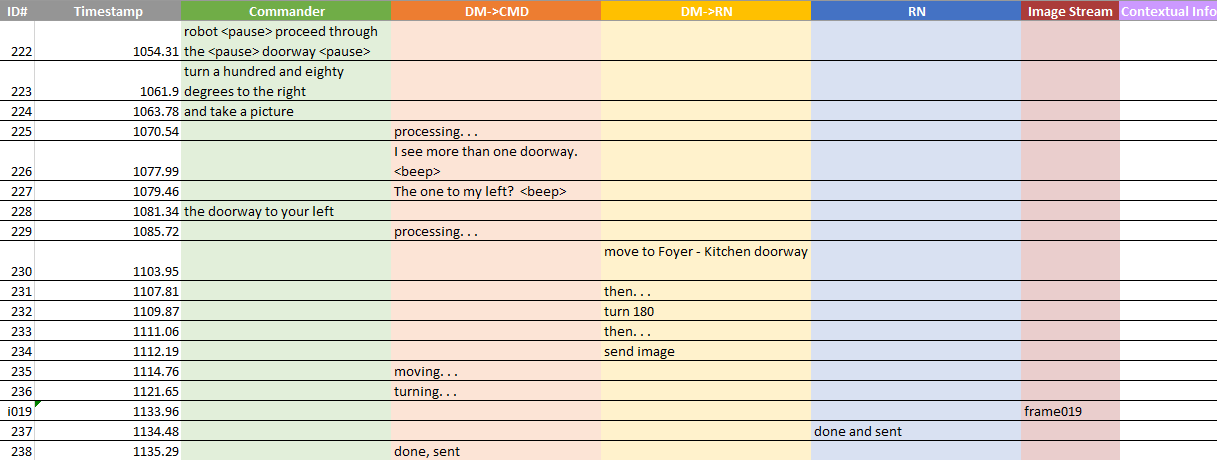}
    \caption{Aligned .xlsx transcript screenshot format with image references}
    \label{fig:iscout}
\end{figure*}

\begin{figure*}[t!]
    \centering
    \includegraphics[width=\textwidth]{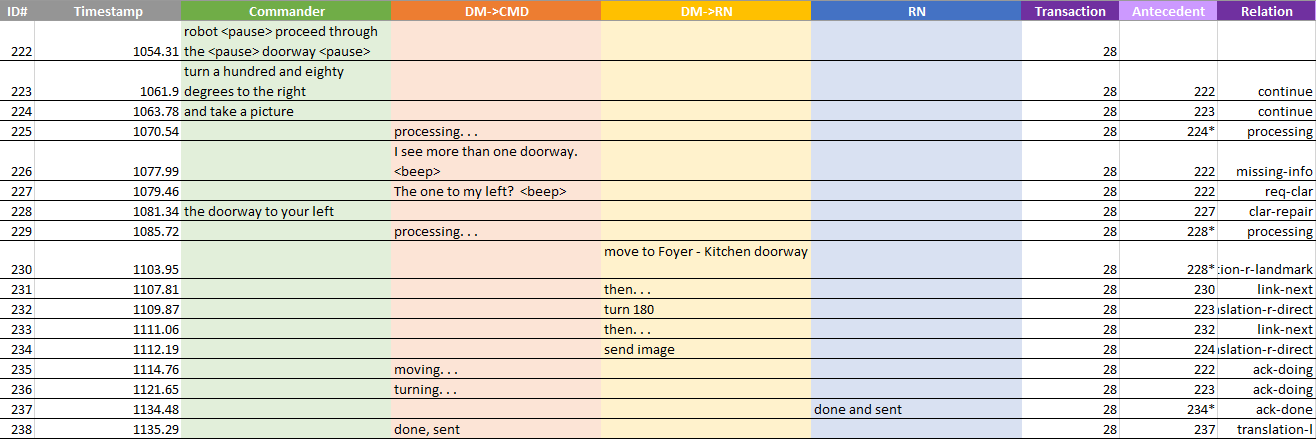}
    \caption{Aligned .xlsx transcript screenshot format with Dialogue Structure annotations (the three rightmost columns)}
    \label{fig:annotations}
\end{figure*}

The corpus statistics from Table~\ref{tab:statistics-main} are further broken up by experiment here in Table~\ref{tab:statistics-extra}. Screenshots of the .xlsx formatted files are shown in Figures~\ref{fig:transcript}-\ref{fig:annotations}.

\begin{table*}[h!]
\begin{footnotesize}
	\centering
	\begin{tabular}{rlrrrrrr}
	\toprule
		 \multicolumn{2}{c}{{\bf Corpus Attribute}} & {\bf Exp. 1} & {\bf Exp. 2} & {\bf Exp. 3 (Subset)} & {\bf Exp. 4} & {\bf Total} \\
            \midrule

           SCOUT Totals & Commanders & 10 & 10 & 63 (10) & 10 & {\bf 93} \\
            & Dialogues & 30 & 30 & 188 (30) & 30 & {\bf 278} \\ 
            & Utterances & 6,097 & 8,421 & 59,859 (8,262) & 14,639 & {\bf 89,056} \\
            & Words -- All & 17,818 & 29,081 & 207,004 (29,441) & 51,194 & {\bf 310,095} \\
            & Words -- Unique & 1,905 & 1,172 & 3,134 (1,252) & 811 & {\bf 4,322} \\
            & Images & 835 & 565 & 3,694 (519) & 691 & {\bf 5,785} \\
            & Avg. Utterance per Dialogue & 203 & 280 & 318 (275) & 487 & {\bf 320} \\
            & Avg. Images per Dialogue & 27.8 & 18.8 & 19.5 (17.3) & 23 & {\bf 20.8} \\
            \midrule

            Annotation Totals & Standard-AMR Sentences & 52 & 212 & 305 & --- & {\bf 569} \\ 
            & Dialogue-AMR Sentences & 52 & 212 & 305 & --- & {\bf 569} \\ 
            & Dialogue Structure TUs & 1,005 & 1,243 & 9,169 (1,216) & 2,246 & {\bf 13,663} \\ 
            & Dialogue Structure Relations & 4,700 & 6,728 & 45,648 (6,406) & 12,354 & {\bf 69,430} \\ 

        \bottomrule
	\end{tabular}
	\caption{SCOUT corpus statistics by experiment \label{tab:statistics-extra}}
\end{footnotesize}
\end{table*}

\end{document}